\documentclass{mn2e}

\usepackage{times}

\begin{document}
\journal{astro-ph/0701113}

\title[Information criteria for astrophysics]
{Information criteria for astrophysical model selection}
\author[Andrew R. Liddle]
{Andrew R. Liddle\\ 
Astronomy Centre, University of Sussex, Brighton BN1 9QH, United
Kingdom\\
Institute for Astronomy, University of Hawai`i, 2680 Woodlawn Drive,
Honolulu, Hawai`i 96822, U.S.A.} 
\maketitle
\begin{abstract}
Model selection is the problem of distinguishing competing models,
perhaps featuring different numbers of parameters. The statistics
literature contains two distinct sets of tools, those based on
information theory such as the Akaike Information Criterion (AIC), and
those on Bayesian inference such as the Bayesian evidence and Bayesian
Information Criterion (BIC).  The Deviance Information Criterion
combines ideas from both heritages; it is readily computed from Monte
Carlo posterior samples and, unlike the AIC and BIC, allows for
parameter degeneracy. I describe the properties of the information
criteria, and as an example compute them from WMAP3 data for several
cosmological models. I find that at present the information theory and
Bayesian approaches give significantly different conclusions from that
data.
\end{abstract}

\section{Introduction}

Although it has been widely recognized only recently, model selection
problems are ubiquitous in astrophysics and cosmology. While parameter
estimation seeks to determine the values of a parameter set chosen by
hand, model selection seeks to distinguish between competing choices
of parameter {\em set}. A considerable body of statistics literature
is devoted to model selection [excellent textbook accounts are given
by Jeffreys 1961, Burnham \& Anderson 2002, MacKay 2003, and Gregory
2005] and its use is widespread throughout many branches of
science. For a non-technical overview of model selection as applied to
cosmology, see Liddle, Mukherjee \& Parkinson (2006a), and for an
overview of techniques and applications see Lasenby \& Hobson (2006).

In general, a \emph{model} is a choice of parameters to be varied and
a prior probability distribution on those parameters.  The goal of
model selection is to balance the quality of fit to observational data
against the complexity, or predictiveness, of the model achieving that
fit. This tension is achieved through model selection statistics,
which attach a number to each model enabling a rank-ordered list to be
drawn up. Typically, the best model is adopted and used for further
inference such as permitted parameter ranges, though the statistics
literature has also seen increasing interest in multi-model inference
combining a number of adequate models (e.g.~Hoeting et al.~1999;
Burnham \& Anderson 2004).

There are two main schools of thought in model selection. Bayesian
inference, particularly as developed by Jeffreys culminating in his
classic textbook (Jeffreys 1961) and by many others since, can assign
probabilities to models as well as to parameter values, and manipulate
these probabilities using rules such as Bayes' theorem.
Information-theoretic methods, pioneered by Akaike (1974) with his
Akaike Information Criterion, instead focus on the Kullback--Leibler
information entropy (Kullback \& Leibler 1951) as a measure of
information lost when a particular model is used in place of the
(unknown) true model. Variants on this latter theme include the
Takeuchi Information Criterion (TIC, Takeuchi 1976), which extends the
AIC by droppinging the assumption that the model set considered
includes the true model. Bayesian statistics include the Bayesian
evidence and an approximation to it known as the Bayesian Information
Criterion (BIC, Schwarz 1978), which, despite the name, does not have
an information-theoretic justification.

Given the plethora of possible statistics, one might despair as to
which to use, especially if they give conflicting results.
Cosmologists, in particular, tend to ally themselves with a Bayesian
methodology, for example the use of Markov Chain Monte Carlo (MCMC)
methods to carry out parameter likelihood analyses, and are therefore
tempted to adopt methods advertised as such. However, even if one were
to side automatically against frequentist approaches, the situation
does not appear that clear cut; Burnham \& Anderson (2004) have argued
that the AIC can be derived in a Bayesian way (and the BIC in a
frequentist one), and that one should not casually dismiss a criterion
soundly grounded in information theory.

Nevertheless, in my view the Bayesian evidence is the preferred tool;
in Bayesian inference it is precisely the quantity which updates the
prior model probability to the posterior model probability, and has an
unambiguous interpretation in these probabilistic terms. The problem
with the evidence is the difficulty in calculating it to the required
accuracy, though the situation there has improved with the development
of the nested sampling algorithm (Skilling 2006) and its
implementation for cosmology in the CosmoNest code (Mukherjee,
Parkinson \& Liddle 2006; Parkinson, Mukherjee \& Liddle 2006). This
paper is principally directed at circumstances where the evidence is
not readily calculable, and a simpler model selection technique is
required.

In this article I describe and apply an additional information
criterion, the Deviance Information Criterion (DIC) of Spiegelhalter
et al.~(2002, henceforth SBCL02), which combines heritage from both
Bayesian methods and information theory.  It has interesting
properties. Firstly, unlike the AIC and BIC it accounts for the
situation, common in astrophysics, where one or more parameters or
combination of parameters is poorly constrained by the data. Secondly,
it is readily calculable from posterior samples, such as those
generated by MCMC methods. It has already been used in astrophysics to
study quasar clustering (Porciani \& Norberg 2006).

\section{Model selection statistics}

\subsection{Bayesian evidence}

The Bayesian evidence, also known as the model likelihood and
sometimes, less accurately, as the marginal likelihood, comes from a
full implementation of Bayesian inference at the model level, and is
the probability of the data given the model. Using Bayes theorem, it
updates the prior model probability to the posterior model
probability. Usually the prior model probabilities are taken as equal,
but quoted results can readily be rescaled to allow for unequal ones
if required (e.g.~Lasenby \& Hobson 2006).  In many circumstances the
evidence can be calculated without simplifying assumptions (though
perhaps with numerical errors). It has now been quite widely applied
in cosmology; see for example Jaffe (1996), Hobson, Bridle \& Lahav
(2002), Saini, Weller \& Bridle (2004), Trotta (2005), Parkinson
et al.~(2006), and Lasenby \& Hobson (2006).

The evidence is given by
\begin{equation}
\label{ev}
E \equiv \int {\cal L}(\theta) \, P(\theta) \, d\theta \,,
\end{equation}
where $\theta$ is the vector of parameters being varied in the model
and $P(\theta)$ is the properly-normalized prior distribution of
those parameters (often chosen to be flat). It is the average value of
the likelihood ${\cal L}$ over the entire model parameter space that
was allowed before the data came in. It rewards a combination of data
fit and model predictiveness. Models which fit the data well and make
narrow predictions are likely to fit well over much of their available
parameter space, giving a high average. Models which fit well for
particular parameter values, but were not very predictive, will fit
poorly in most of their parameter space driving the average
down. Models which cannot fit the data well will do poorly in any
event. 

The integral in equation (\ref{ev}) may however be difficult to
calculate, as it may have too many dimensions to be amenable to
evaluation by gridding, and the simplest MCMC methods such as
Metropolis--Hastings produce samples only in the part of parameter
space where the posterior probability is high rather than throughout
the prior. Nevertheless, many methods exist (e.g.~Gregory 2005; Trotta
2005), and the nested sampling algorithm (Skilling 2006) has proven
feasible for many cosmology applications (Mukherjee et al.~2006;
Parkinson et al.~2006; Liddle et al.~2006b).

A particular property of the evidence worth noting is that it does
\emph{not} penalize parameters (or, more generally, degenerate
parameter combinations) which are unconstrained by the data. If the
likelihood is flat or nearly flat in a particular direction, it simply
factorizes out of the evidence integral leaving it unchanged. This is
an appealing property, as it indicates that the model fitting the data
is doing so really by varying fewer parameters than at first seemed to
be the case, and it is the unnecessary parameters that should be
discarded, not the entire model.

\subsection{AIC and BIC}

Much of the literature, both in astrophysics and elsewhere, seeks a
simpler surrogate for the evidence which still encodes the tension
between fit and model complexity. In Liddle (2004), I described two
such statistics, the AIC and BIC, which have subsequently been quite
widely applied to astrophysics problems. They are relatively simple to
apply because they require only the maximum likelihood achievable
within a given model, rather than the likelihood throughout the
parameter space. Of course, such simplification comes at a cost, the
cost being that they are derived using various assumptions,
particularly gaussianity or near-gaussianity of the posterior
distribution, that may be poorly respected in real-world situations.

The AIC is defined as
\begin{equation}
\mathrm{AIC} \equiv -2\ln {\cal L}_{{\rm max}} + 2k\,,
\end{equation}
where ${\cal L}_{{\rm max}}$ is the maximum likelihood achievable by
the model and $k$ the number of parameters of the model (Akaike
1974). The best model is the one which minimizes the AIC, and there
is no requirement for the models to be nested.  The AIC is derived by
an approximate minimization of the Kullback--Leibler information
entropy, which measures the difference between the true data
distribution and the model distribution.  An explanation geared to
astronomers can be found in Takeuchi (2000), while the full
statistical justification is given by Burnham \& Anderson (2002).

The BIC was introduced by Schwarz (1978), and is defined as
\begin{equation}
\mathrm{BIC} \equiv -2\ln{\cal L}_{{\rm max}} + k \ln N \,,
\end{equation}
where $N$ is the number of datapoints used in the fit.  It comes from
approximating the evidence ratios of models, known as the Bayes factor
(Jeffreys 1961; Kass \& Raftery 1995). The BIC assumes that the
datapoints are independent and identically distributed, which may or
may not be valid depending on the dataset under consideration (e.g.~it
is unlikely to be good for cosmic microwave anisotropy data, but may
well be for supernova luminosity-distance data).

Applications of these two criteria have usually shown broad agreement
in the conclusions reached, but occasional differences in the detailed
ranking of models. One should consider the extent to which the
conditions used in the derivation of the criteria are violated in real
situations. A particular case in point is the existence of parameter
degeneracies; inclusion (inadvertent or otherwise) of unconstrained
parameters is penalized by the AIC and BIC, but not by the
evidence. Interpretation of the BIC as an estimator of evidence
differences is therefore suspect in such cases.

Burnham \& Anderson (2002, 2004) have stressed the importance of
using a version of the AIC corrected for small sample sizes, AIC$_{\rm
c}$. This is given by (Sugiura 1978)
\begin{equation}
\mathrm{AIC}_{\rm c} = \mathrm{AIC} + \frac{2k(k+1)}{N-k-1}\,.
\end{equation}
Since the correction term anyway disappears for large sample sizes, $N
\gg k$, there is no reason not to use it even in that case, i.e.~it is
\emph{always} preferable to use AIC$_{\rm c}$ rather than the original
AIC. In typical small-sample cases, e.g.~$N/k$ being only a few, the
correction term strengthens the penalty, bringing the AIC$_{\rm c}$
towards the BIC and potentially mitigating the difference between
them.

\subsection{DIC}

The DIC was introduced by SBCL02. It has already been widely applied
outside of astrophysics. Its starting point is a definition of an
effective number of parameters $p_D$ of a model. This quantity, known
also as the \emph{Bayesian complexity}, has already been introduced
into astrophysics by Kunz, Trotta \& Parkinson (2006), with focus on
assessing the number of parameters that can be usefully constrained by
a particular dataset.

It is defined by
\begin{equation}
p_D = \overline{D(\theta)}-D(\bar{\theta}), \quad \mathrm{where} \;
D(\theta) = -2 \ln {\cal L}(\theta) + C \,.
\end{equation}
Here $C$ is a `standardizing' constant depending only on the data
which will vanish from any derived quantity, and $D$ is the deviance
of the likelihood. The bars indicate averages over the
posterior distribution. In words, then, $p_D$ is the mean of
the deviance, minus the deviance of the mean.
If we define an effective chi-squared as usual by $\chi^2 = -2\ln
{\cal L}$, we can write
\begin{equation}
p_D=\overline{\chi^2(\theta)} - \chi^2(\bar{\theta}) \,.
\end{equation}

Its intent becomes clear from studying a simple one-dimensional
example, in which the likelihood is a gaussian of zero mean and width
$\sigma$, i.e.~$\ln {\cal L} = A-x^2/2\sigma^2$, and where the prior
distribution is flat with width $a \sigma$. Care is needed to properly
normalize the posterior, which relates the likelihood amplitude $A$ to
the prior width.  In the limit where $a \gg 1$, so that the posterior
is well confined within the prior, one finds $p_D = 1$ (in this case,
the averaging is just evaluating the variance of the distribution, but
in units of that variance). This corresponds to a well-measured
parameter. If instead $a \ll 1$, so that the data are unable to
constrain the parameter, then $p_D \rightarrow 0$ since $\chi^2$
becomes independent of $x$. Hence $p_D$ indicates the number of
parameters actually constrained by the data. Extension of the above
argument to an $N$-dimensional gaussian, potentially with covariance,
indicates $p_D = N$ if all dimensions are well contained within the
prior, and $p_D < N$ otherwise (SBCL02; Kunz et al.~2006).

One issue of debate in the statistics literature is the choice of the
mean parameter value in the definition of $p_D$. One could
alternatively argue for the maximum likelihood in its place. This
choice affects the possible reparametrization dependence of the
statistic (SBCL02; Celeux et al.~2006). It may be that the best choice
depends on the situation under study (e.g.~the mean parameter value
will be a poor choice if the likelihood has distinct strong peaks).

The DIC is then defined as
\begin{equation}
\mathrm{DIC}  \equiv  D(\bar{\theta})+2p_D = \overline{D(\theta)}+p_D \,.
\end{equation}
The first expression is motivated by the form of the AIC, replacing
the maximum likelihood with the mean parameter likelihood, and the
number of parameters with the effective number. It can therefore be
justified on information/decision theory grounds, as discussed by
SBCL02. The second form is interesting because the mean deviance can be
justified in Bayesian terms, which always deal with model-averaged
quantities rather than maximum values.

The DIC has two attractive properties:
\begin{enumerate}
\item It is determined by quantities readily obtained from Monte Carlo
  posterior samples. One simply averages the deviances over the
  samples. If the calculation is being done by whoever generated the
  chains, they can obtain the deviance at the mean with a single extra
  likelihood call, but even if using chains generated by others, it
  should be fine to use the sample closest to that mean value as the
  estimator, especially bearing in mind the possibility that the mode
  could have been used in place of the mean. The calculation is also
  easily done with posterior samples generated by nested sampling,
  which have non-integer weights (Parkinson et al.~2006).
\item By using the effective number of parameters, the DIC overcomes
  the problem of the AIC and BIC that they do not discount parameters
  which are unconstrained by the data. 
\end{enumerate}

Note that in the case of well-constrained parameters, the DIC
approaches the AIC and not the BIC, since $D(\bar{\theta}) \rightarrow
-2\ln {\cal L}_{{\rm max}}$ and \mbox{$p_D \rightarrow k$}. It is
plausible to believe that it too can be corrected for small dataset
sizes using the same formula that leads to AIC$_{\rm c}$, though to my
knowledge there is currently no proof of this.

\subsection{Other criteria}

In addition to those already mentioned, the literature contains many
other information criteria, but mostly sharing the heritage of those
above. The TIC (Takeuchi 1976) generalizes the AIC by dropping the
assumption that the true model is in the set considered, but in
practice is hard to compute and, where computation has been carried
out, tends to give results very similar to the AIC (Burnham \&
Anderson 2002, 2004). A Bayesian version of the AIC, the Expected AIC
(EAIC), where one takes its expected value over the posterior
distribution rather than evaluating at the maximum, has been proposed
(by Brooks in the comments to SBCL02) but does not appear to have been
significantly applied.

Other information criteria, which appear to have been less
widely used, include the Network Information Criterion (NIC), the
Subspace Information Criterion (SIC, though this abbreviation is
sometimes used for Schwarz Information Criterion as another name for
the BIC), and the Generalized Information Criterion (GIC). The DIC
also comes in many variants, see e.g.~Celeux et al.~(2006).

An interesting variant was proposed by Sorkin (1983), using a Turing
machine construction to define an entropy associated with the theory
to be used as a penalty term. This was recently applied to
cosmological data by Magueijo \& Sorkin (2006). It has not been picked
up by the statistics community, but may be related to the widely-used
minimum message length paradigm (Wallace \& Boulton 1968; Wallace
2005). The idea of interpetting the best model as the one offering
maximal algorithmic compression of the data goes all the way back to
late 17th century writings by Leibniz.

\begin{table*}
\caption{\label{tab1} Results for comparison of different models to
WMAP3 data. The differences are quoted with respect to the first
model. Negative is preferred.}
\begin{tabular}{lcccccccc}
Model & Parameters $k$  & $p_D$ & $-2\ln {\cal L}(\bar{\theta})$ & DIC
  &  $-2\ln {\cal L}_{{\rm max}}$ & $\Delta$DIC & $\Delta$AIC$_c$ &
  $\Delta$BIC\\  
\hline
Base+$A_{{\rm SZ}}$ & 6 & 5.2 & 11262.6  & 11272.9 & 11262.2 & 0
  & 0 & 0 \\ 
Base+$n_{{\rm S}}$ & 6 & 6.3 & 11253.3  & 11265.9 & 11252.5 & -7.0 &
  -9.7 & -9.7\\  
Base+$A_{{\rm SZ}}$+$n_{{\rm S}}$ & 7 & 5.6 & 11253.0 & 11264.1 &
  11252.6 & -8.8 & -7.6 & -2.3\\ 
Base+$A_{{\rm SZ}}$+$n_{{\rm S}}$+$r$ & 8 & 5.4 & 11254.2  & 11265.0 &
  11252.6 & -7.9 & -5.6 & +5.0\\ 
Base+$A_{{\rm SZ}}$+$n_{{\rm S}}$+running & 8 & 6.2 & 11250.0  &
  11262.3 & 11249.0 & -10.6 & -9.2 & +1.4
\end{tabular}
\end{table*}

\subsection{Dimensional consistency and model selection philosophy}

Dimensional consistency refers to the behaviour of the model selection
statistics in the limit of arbitrarily large datasets. The BIC and
evidence are dimensionally consistent, meaning that if one of the
considered models is true, they give 100 per cent support to that
model as the dataset becomes large. As a necessary consequence,
however, they will give 100 per cent support to the best model even if
it is not true. By contrast, the AIC is dimensionally inconsistent
(Kashyap 1980), sharing its support around the models even with
infinite data. As the DIC approaches the AIC in the limit of large
datasets, it too is dimensionally inconsistent (SBCL02).

Dimensional consistency does not seem to particularly bother most
statisticians, as they are typically seeking models which can explain
data and have some predictive power, rather than expecting to
represent some underlying truth. Indeed, they commonly quote
statistician George Box: ``All models are wrong, but some are
useful.''  The problem of dimensional consistency is therefore
mitigated, because they do not expect the set of models to remain
static as the dataset evolves.  Cosmologists, however, are probably
not yet willing to concede that they might be looking for something
other than absolute truth specified by a finite number of
parameters. Combining this line of argument with statements above,
this implies that the Bayesian evidence indeed is the preferred choice
for cosmological model selection when it can be calculated.

\section{Information criteria for WMAP3}

I now apply the information criteria to WMAP3 model fits as compiled
by the WMAP team on LAMBDA.\footnote{Legacy Archive for Microwave
Background Data Analysis: http://lambda.gsfc.nasa.gov. Chains were
downloaded in December 2006. The subsequent January 2007 update does
not allow model selection as the chains were not all generated with
the same likelihood code.} The DIC calculation is straightforward. The
8 chains for each cosmology are concatenated, the mean deviance found
by averaging the likelihoods, and the deviance at the mean
estimated by finding the MCMC point located closest to the mean (where
the distance in each parameter direction was measured in units of the
standard deviation of that parameter).

I also quote the values of the differences in AIC$_c$ and BIC, where
the maximum likelihood is taken directly from the most likely
posterior sample (in principle this may slightly disadvantage models
with more parameters, for which the most likely sample will typically
be slightly further from the true maximum, though for the WMAP3 sample
sizes this effect will be small). I take $N$ to
be the number of power spectrum datapoints, $N_{{\rm WMAP3}}=1448$
(Spergel et al.~2006), this choice to be discussed further below
(nothing changes significantly if a slightly larger number $\sim 3000$
is used to allow for the pixel-based treatment of the low-$\ell$
likelihood). With this large value, $\Delta$AIC and $\Delta$AIC$_c$ are
indistinguishable.

The available model fits unfortunately do not quite cover all cases
that might be of interest. All well-fitting models vary five standard
parameters, being the physical baryon density $\Omega_{{\rm b}} h^2$,
the physical cold dark matter density $\Omega_{{\rm c}} h^2$, the
sound horizon $\theta$, the perturbation amplitude $\ln(10^{10}
A_{{\rm S}})$, and the optical depth $\tau$ (the Hubble constant and
dark energy density are derived parameters). However fits varying just
these parameters, a Harrison--Zel'dovich model suggested as the best
model from first-year WMAP data in Liddle (2004), are not
available. Nevertheless, I will refer to this as the Base model.
Instead, there are two different six-parameter models, one adding the
spectral index $n_{{\rm S}}$ and one adding the phenomenological
Sunyaev--Zel'dovich (SZ) marginalization parameter $A_{{\rm SZ}}$
(Spergel et al.~2006).  All further available models include $A_{{\rm
SZ}}$; extra parameters that I then consider are the spectral index
$n_{{\rm S}}$ (giving the standard $\Lambda$CDM model), further
addition of tensors $r$ to give the standard slow-roll inflation
model, and inclusion of spectral index running (without tensors).

The main subtlety is the inclusion of $A_{{\rm SZ}}$. This is poorly
constrained by the data and hence is not expected to contribute fully
to $p_D$; nevertheless the likelihood does have some dependence on it
and it must be included in the analysis that determines the deviance
at the mean. Of the parameters considered, $A_{{\rm SZ}}$ and $\tau$
are phenomenological parameters which, at least in principle though
not yet in practice, can be determined from the others. The remaining
four are truly independent according to present understanding.

The uncertainty in the DIC may not be well estimated by analyzing
subsamples, as with smaller samples the mean deviance will be less
well estimated by the nearest point. Instead I estimated the
uncertainty by employing bootstrap resamples of the combined sample
list. This showed that the statistical accuracy was limited by the
accuracy with which the $\ln {\cal L}$ values were stored, $\pm 0.1$
corresponding to $\pm 0.2$ in the DIC. As this is a much smaller
uncertainty than the level at which differences are significant, 
the statistical uncertainty in the determination of the DIC
is negligible.

The results are shown in Table~\ref{tab1}. The $p_D$ values are in
good agreement with expectation. Kunz et al.~(2006) computed $p_D$ for
several models using a compilation of microwave anisotropy data
including WMAP3, and always found $p_D$ close to the input number of
parameters. However they ran their own chains and did not include the
poorly-constrained parameters $A_{{\rm SZ}}$ and $r$. Models including
those parameters return a $p_D$ significantly less than $k$.

While only the Bayesian evidence has the full interpretation as the
model likelihood, leading to the posterior model probability, the AIC
has also been interpretted as a model likelihood by defining Akaike
weights (Akaike 1981; Burnham \& Anderson 2004)
\begin{equation}
w_i = \frac{\exp(-\Delta {\rm AIC}_{c,i}/2)}{\sum_{r=1}^R \exp (-\Delta
    {\rm AIC}_{c,r}/2)} \,,
\end{equation}
where there are $R$ models and the differences are with respect to any
one. The same interpretation can be given to the DIC differences
(SBCL02). For the BIC, insofar as it well approximates twice the log
of the Bayes factor, it too can be interpreted as a model
likelihood. By convention significance is then judged on the Jeffreys'
scale, which rates $\Delta {\rm IC} > 5$ as `strong' and $\Delta {\rm
IC}>10$ as `decisive' evidence against the model with higher criterion
value. If the interpretation as model likelihoods holds, these points
correspond to odds ratios of approximately 13:1 and 150:1 against the
weaker model. As with the evidence, these likelihoods can be further
weighted by a prior model probability if desired.

Recall that the DIC, like the AIC, is motivated from information
theory, while the BIC is not. Indeed, we see that the DIC results
quite closely follow the AIC results; both argue quite strongly
against the Base+$A_{{\rm SZ}}$ model, but are then rather
inconclusive amongst the remaining models. So information theory
methods are neither for nor against inclusion of extra parameters such
as $r$ and running at this stage.  Incidentally, we can also see that
if the DIC were defined using ${\cal L}_{{\rm max}}$ rather than
${\cal L}(\bar{\theta})$, little difference would have arisen in this
comparison.

The information criteria indicate that WMAP3 has put the
Harrison--Zel'dovich model (with SZ marginalization) under
considerable, if not yet conclusive, pressure. This is in accord with
the conclusions reached by Spergel et al.~(2006) using chi-squared per
degree of freedom arguments, though the information criterion give
weaker support to this conclusion by recognizing model
dimensionality. The strength of conclusion against
Harrison--Zel'dovich could also be weakened by various systematic
effects in data analysis choices, e.g.~inclusion of gravitational
lensing (Lewis 2006), beam modelling (Peiris \& Easther 2006), and
point-source subtraction (Eriksen et al.~2006; Huffenberger, Eriksen
\& Hansen 2006).

By contrast, Bayesian approaches do not put $n_{{\rm S}}=1$ under any
kind of pressure. Parkinson et al.~(2006) found that the full
evidences for the Base model and Base+$n_{{\rm S}}$ were
indistinguishable with WMAP3 alone, and still inconclusive with
inclusion of other datasets. However that analysis did not include SZ
marginalization, and so the equivalent comparison cannot be made
here. However the BIC comparison between those models each with
$A_{{\rm SZ}}$ added does not show any strong preference, and it seems
a safe bet that had the Base model itself been supplied by WMAP3, its
BIC difference compared with Base+$n_{{\rm S}}$, the best model in the
set as judged by the BIC, would not have been significant.

Further, while the information theory methods are ambivalent about $r$
and running, the BIC argues rather strongly against, especially in the
case of tensors which offer no improvement at all in
data-fitting. Full evidence calculations however show that this
conclusion is quite prior dependent (Parkinson et al.~2006).

That the two methods give such different answers is due to the way
that prior assumptions are treated, in particular the prior widths of
the parameter ranges. The AIC does not care about this at all, and the
DIC only cares while the data is weak enough that some prior
information on the parameter distribution remains. By contrast, in 
Bayesian model comparison the prior width is a key concept,
determining the predictiveness of the model. For the evidence this is
reflected in the domain of integration over which the likelihood is
averaged, while for the BIC it is in the dependence on the amount of
data. Cosmologists are in the fortunate position that for many
parameters the likelihood is highly compressed within reasonable
priors, forcing a discrepancy between information theory and Bayesian
results. This discrepancy will be further enhanced in the future if
the data continue to improve without requiring evolution in the model
dataset, i.e.~the problem of dimensional inconsistency of the AIC/DIC
may already be with us.

Concerning the inclusion of $A_{{\rm SZ}}$ in models, it is clear that
Bayesian methods don't like including it as a fit parameter, since it
is poorly constrained and does not significantly improve the fit.
However the SZ effect is certainly predicted to be in the data at some
level, though it ought to be derived from the other parameters rather
than fit. It is tempting to try to deal with this by using $p_D$ in
the BIC rather than $k$, but there is no existing justification for
doing so. The same issue does not arise with the optical depth, also a
derived parameter, as it is well constrained by the data in all
models.

In computing the BIC above, I adopted the number of datapoints
literally. This may not always be the best choice: the derivation of
the BIC requires the data to be independent and identically
distributed, and it may be that this can be better achieved by binning
the data in some suitable way. However to do so would require a whole
new likelihood analysis for the binned data, counter to the desire
here that the methods should be applicable to pre-existing posterior
samples. In any case there does not appear to be any well-defined way
to judge how much binning, if any, is desirable.

Finally, I note that while here it is the BIC which appears to behave
most like the evidence, in their quasar clustering studies Porciani \&
Norberg (2006) found that the DIC was the only criterion to give
precisely the same model ranking order and level of inconclusiveness
as the Bayes factors, with the BIC underfitting.

\section{Summary}

I have described several information criteria that can be used for
astrophysical model selection, representing the rival strands of
information theory and Bayesian inference. In application to WMAP3
data, the DIC behaves rather similarly to the AIC, despite the
presence of parameter degeneracies. The conclusions one would draw
from those statistics are rather different from those indicated by
Bayesian methods, either the full evidence as computed in Parkinson et
al.~(2006) or the BIC as calculated in this article. 

\section*{Acknowledgments}

This research was supported by PPARC. Thanks to David Parkinson and
Pia Mukherjee for advice on analyzing WMAP chains, and Istvan Szapudi
for discussions.

\bsp


\begin{thebibliography}{}
\bibitem[Akaike 1974]{Akaike} Akaike H., 1974, IEEE T. Automat. Contr., 
	19, 716
\bibitem[Akaike 1981]{Akaike2} Akaike H., 1981, J. Econometrics, 16, 3
\bibitem[Burnham \& Anderson 2002]{BA02} Burnham K. P., Anderson
	D. R., 2002, \emph{Model selection and multimodel inference},
	2nd ed., Springer-Verlag, New York
\bibitem[Burnham \& Anderson 2004]{BA04} Burnham K. P., Anderson
        D. R., 2004, Sociol. Method. Res., 33, 261 
\bibitem[Celeux et al]{Cetal} Celeux G., Forbes F., Robert C. P.,
  Titterington D. M., 2006, Bayesian Anal., 1, 651
\bibitem[Eriksen et al. 2006]{Eetal} Eriksen H. K. et al., 2006,
  astro-ph/0606088 
\bibitem[Gregory 2005]{gregory} Gregory P., 2005, \emph{Bayesian
  logical data analysis for the physical sciences}, Cambridge
  University Press 
\bibitem[Hobson, Bridle \& Lahav]{HBL} Hobson M. P., Bridle S. L.,
  Lahav O., 2002, MNRAS, 335, 377
\bibitem[Hoeting et al.~1999]{HMRV} Hoeting J. A., Madigan D., Raftery
  A. E., Volinsky C. T., 1999, Stat. Science, 14, 382
\bibitem[Huffenberger et al.~2006]{Hetal} Huffenberger K. M., Eriksen
  H. K., Hansen F. K., 2006, ApJL, 651, L81
\bibitem[Jaffe 1996]{Jaff} Jaffe A., 1996, ApJ, 471, 24
\bibitem[Jeffreys 1961]{Jeff} Jeffreys H., 1961, \emph{Theory of
	probability}, 3rd ed., Oxford University Press
\bibitem[Kashyap 1980]{Kashyap} Kashyap R., 1980, IEEE
	T. Automat. Contr., 25, 996
\bibitem[Kass \& Raftery 1995]{KR95} Kass R. E., Raftery A. E., 1995,
	J. Am. Stat. Assoc., 90, 773
\bibitem[Kullback \& Leibler 1951]{KL51} Kullback S., Leibler R. A.,
  1951, Ann. Math. Stat., 22, 79
\bibitem[Kunz, Trotta \& Parkinson 2006]{KTP} Kunz M., Trotta R.,
  Parkinson D., 2006, Phys. Rev. D, 74, 023503
\bibitem[Lasenby \& Hobson 2006]{LH} Lasenby A., Hobson M. P.,
  Proc. Science (CMB2006) 014
\bibitem[Lewis 2006]{Lew06} Lewis A., 2006, astro-ph/0603753
\bibitem[Liddle 2004]{Lid04} Liddle A. R., 2004, MNRAS, 351, L49
\bibitem[Liddle, Mukherjee \& Parkinson 2006a]{LMP} Liddle A. R.,
  Mukherjee P., Parkinson D., 2006a, A\&G, 47, 4.30 
\bibitem[Liddle, et al.~2006b]{LMPW} Liddle A. R.,
  Mukherjee P., Parkinson D., Wang Y., 2006b, Phys. Rev. D, 74,
  123506 
\bibitem[MacKay 2003]{mackay} MacKay D. J. C., 2003, \emph{Information
	theory, inference, and learning algorithms}, Cambridge
	University Press
\bibitem[Magueijo \& Sorkin 2006]{MS06} Magueijo J., Sorkin R. D.,
  2006, astro-ph/0604410
\bibitem[Mukherjee, Parkinson \& Liddle 2006]{MPL} Mukherjee P.,
  Parkinson D., Liddle A. R., 2006, ApJL, 638, L51
\bibitem[Parkinson, Mukherjee \& Liddle 2006]{PML} Parkinson D.,
  Mukherjee P., Liddle A. R., 2006, Phys. Rev. D, 73, 123523, code at
  http://cosmonest.org
\bibitem[Peiris \& Easther 2006]{PE06} Peiris H., Easther R., 2006,
  JCAP, 0607, 002
\bibitem[Porciani \& Norberg 2006]{PN06} Porciani C., Norberg P.,
  2006, MNRAS, 371, 1824
\bibitem[Saini, Weller \& Bridle 2004]{SWB} Saini T. D., Weller J.,
  Bridle S. L., 2004, MNRAS, 348, 603
\bibitem[Schwarz 1978]{Schwarz} Schwarz G., 1978, Ann. Statist., 
	5, 461
\bibitem[Skilling 2006]{Skill06} Skilling J., 2006, Bayesian Anal.,
  1, 833  
\bibitem[Sorkin 1983]{Sor} Sorkin R. D., 1983, Int. J. Theor. Phys.,
  22, 1091
\bibitem[Spergel et al.~2006]{Sper} Spergel D. N. et al.~(the WMAP
	Team), 2006, astro-ph/0603449
\bibitem[Spiegelhalter et al]{S} Spiegelhalter D. J., Best N. G.,
  Carlin B. P., van der Linde A., 2002, J. R. Statist. Soc. B, 64, 583
  [SBCL02] 
\bibitem[Sugiura 1978]{Sug78} Sugiura N., 1978,
  Commun. Stat. A-Theor., A7, 13   
\bibitem[Takeuchi 1976]{Tak76} Takeuchi K., 1976, Suri-Kagaku
  (Math. Sci.) 153, 12 [in Japanese]
\bibitem[Takeuchi 2000]{Tak00} Takeuchi T. T., 2000, Astrophys. Space
	Sci., 271, 213
\bibitem[Trotta 2005]{Trot} Trotta R., 2005, astro-ph/0504022
\bibitem[Wallace 2005]{wall} Wallace C. S., 2005, \emph{Statistical
  and Inductive Inference by Minimum Message Length}, Springer
\bibitem[Wallace \& Boulton 1968]{wallboul} Wallace C. S., Boulton D. M.,
  1968, Comput. J., 11, 185
\end{thebibliography}
\end{document}